\documentclass[pra,aps,amsmath,amssymb,showkeys,showpacs]{revtex4}
\usepackage{graphicx}
\newcommand{\mm}{\mathsf{M}}

\begin{document}
\clearpage
\preprint{}

\title{On conclusive eavesdropping and measures of mutual information in quantum key distribution}

\author{Alexey E. Rastegin}

\affiliation{
Department of Theoretical Physics, Irkutsk State University,
Gagarin Bv. 20, Irkutsk 664003, Russia
}

\begin{abstract}
We address the question of quantifying eavesdropper's information
gain in an individual attack on systems of quantum key
distribution. It is connected with the concept of conclusive
eavesdropping introduced by Brandt. Using the BB84 protocol, we
examine the problem of estimating a performance of conclusive
entangling probe. The question of interest depends on the choice
of a quantitative measure of eavesdropper's information about the
error-free sifted bits. The Fuchs--Peres--Brandt probe realizes a
very powerful individual attack on the BB84 scheme. In the usual
formulation, Eve utilizes the Helstrom scheme in distinguishing
between the two output probe states. In conclusive eavesdropping,
the unambiguous discrimination is used. Comparing these two
versions allows to demonstrate serious distinctions between widely
used quantifiers of mutual information. In particular, the
so-called R\'{e}nyi mutual information does not seem to be a
completely legitimate measure of an amount of mutual information.
It is brightly emphasized with the example of conclusive
eavesdropping. 
\end{abstract}

\maketitle

\pagenumbering{arabic}
\setcounter{page}{1}

\section{Introduction}\label{sec1}

In last years of his scientific activity, Howard Brandt paid a
significant attention to questions of quantum cryptography.
Quantum cryptography provides a long-term solution to the problem
of communication security
\cite{lomonaco,tittel,assche06,barnett09}. In a series of papers
\cite{brandt03q,brandt05pra,brandt05,brandt05b,brandt05ms,brandt06jmo,brandt06m},
Brandt addressed the problem of optimizing eavesdropper's probe
for attacking the BB84 protocol of quantum key distribution. The
BB84 protocol \cite{bb84} is probably the most known scheme of
quantum key distribution. One of approaches to eavesdropping is
connected with unambiguous state discrimination \cite{brandt05}.
In a version of the B92 scheme, the two carriers are distinguished
by the POVM for unambiguous state discrimination
\cite{lomonaco,palma}. A photonic implementation of such POVMs is
considered in \cite{huttner96,brandt99}. The papers
\cite{brandt05,brandt05b,brandt05ms} addressed the case, when Eve
tries to get data from her probe by means of unambiguous
discrimination. It must be stressed that an efficiency of
eavesdropping also depends on the used reconciliation method
\cite{herb08}.

As was discussed in \cite{brandt05}, the BB84 protocol may
sometimes have a vulnerability analogous to the well-known
vulnerability of the B92 protocol. The latter has further been
considered in \cite{shapiro06}. Here, the principal point concerns
the effect of the inconclusive answer. To take it into account
correctly, we should choose a proper information-theoretic
measure. Using the so-called ``R\'{e}nyi mutual information'', the
author of \cite{shapiro06} found a certain weakness of the attack
modification with unambiguous discrimination in comparison with
the usual one. We will show that the amount of this
weakness is essentially token in character. Indeed, it is
considerably determined by the chosen entropic order in the
measure of mutual information. The made comparison provides an
interesting example, which helps to reveal some inadequacy of
the widely used measure of mutual information. Since it
reflects existing distinctions in information-theoretic
properties, we also need in a detailed discussion of
used quantities.

The contribution of this paper is two-fold. First, we aim to
justify that the standard mutual information seems to be a
preferred measure in estimating a performance of eavesdropper's
probe. At least, it is better in studies of conclusive
eavesdropping. Second, we motivate that the modified attack with
unambiguous state discrimination is still of interest and deserves
further investigations. Security of systems for quantum key
distribution against eavesdropping is a difficult problem with
many facets. Security analysis should take into account many
scenarios possible in practice. To understand various aspects
deeper, we prefer to consider them separately. In the present
paper, we will focus on possible ways to quantify an information
obtained by eavesdropper during an individual attack. Using
inappropriate measure of information may lead to a wrong opinion
about an essential weakness of conclusive eavesdropping.

The paper is organized as follows. In Section \ref{sec2}, we
discuss entropies and related information-theoretic terms.
Considered properties are of great importance in interpretation of
measures of mutual information. So, the property to show a
reduction in the uncertainty of one random variable due to 
knowledge of other is based on the chain rule for the conditional
entropy. In Section \ref{sec3}, we briefly describe two versions
of the Fuchs--Peres--Brandt (FPB) probe for eavesdropping the BB84
scheme. Main findings of this paper are presented in Section
\ref{sec4}. Using the two versions of the FPB probe, we compare
their performance from the viewpoint of three measures of mutual
information. In Section \ref{sec5}, we conclude the paper with a
summary of results.

\section{Entropies and related notions of information theory}\label{sec2}

In this section, we review some material from information theory.
Quantum key distribution is a procedure used by Alice and Bob for
obtaining two identical copies of a random and secret sequence of
bits. Making some intrusion into a communication channel, Eve try
to learn original bits. During this process, each of the three
parties will obtain some string of bits. The three strings can be
interpreted as binary random variables \cite{palma}. A degree of
dependence between two random variables is typically measured in
terms of the mutual information.

We begin with basic entropic functions. Let discrete random
variable $X$ take values on finite set with the probability
distribution $\{p(x)\}$. The Shannon entropy of $X$ is defined as
\cite{CT91}
\begin{equation}
H(X):=-\sum\nolimits_{x}{p(x)\,\log{p}(x)}
\ . \label{shnent}
\end{equation}
As a rule, the range of summation will be clear from the context.
The logarithm in (\ref{shnent}) is taken to the base $2$. Let $Y$
be another random variable, and let $p(x,y)$ denote corresponding
joint probabilities. The joint entropy $H(X,Y)$ is defined like
(\ref{shnent}) by substituting the joint probabilities. An
adequacy of the quantity (\ref{shnent}) in applications to quantum
measurement statistics was discussed by Brukner and Zeilinger
\cite{brz01}. Further development of their approach to information
is considered in \cite{bzail15}.

Dealing with the notion of conditional entropy, we introduce the
function
\begin{equation}
H(X|y):=-\sum\nolimits_{x}{p(x|y)\,\log{p}(x|y)}
\ , \label{pshen}
\end{equation}
where $p(x|y)=p(x,y)/p(y)$. Then the entropy of $X$ conditional on
knowing $Y$ is defined as \cite{CT91}
\begin{equation}
H(X|Y):=\sum\nolimits_{y}{p(y){\,}H(X|y)}
\ . \label{cshen}
\end{equation}
The entropy is concave in probabilities,
as the function $\xi\mapsto-\xi\log\xi$ is concave. It follows
from concavity that \cite{CT91}
\begin{equation}
H(X|Y,Z)\leq{H}(X|Y)
\ . \label{coonm}
\end{equation}
In other words, conditioning on more reduces entropy.
This property is essential for a treatment of (\ref{cshen}) just
as the conditional entropy. The definition (\ref{cshen}) is
further motivated by the chain rule \cite{CT91}
\begin{equation}
H(X,Y)=H(X|Y)+H(Y)=H(Y|X)+H(X)
\ . \label{schr}
\end{equation}
It directly gives $H(X|Y)=H(X,Y)-H(Y)$, whence we see remaining
lack of knowledge about $X$ at the given $Y$.

The notion of mutual information aims to measure how much
information $X$ and $Y$ have in common \cite{nielsen}. The
definition of mutual information is expressed as \cite{CT91}
\begin{equation}
I(X,Y):=H(X)+H(Y)-H(X,Y)
\ . \label{mindf}
\end{equation}
It is clearly symmetric in entries. By the chain rule
(\ref{schr}), we can also rewrite (\ref{mindf}) as
\begin{equation}
I(X,Y)=H(X)-H(X|Y)=H(Y)-H(Y|X)
\ . \label{dfmin}
\end{equation}
So, the mutual information shows a reduction in the uncertainty of
one random variable due to knowledge of other \cite{CT91}. In the
context of quantum cryptography, the mutual information was first
applied in \cite{palma}. Such an approach was motivated with using
some ideas of the paper \cite{csk78}. In the following, we will
discuss some extensions of the above quantities to generalized
entropic functions. Note that such generalizations typically lose
some of the essential properties used in information theory. The
mutual information (\ref{mindf}) is also used in definition of the
Shannon distinguishability \cite{graaf,biham}. Fuchs gave a
comprehensive presentation of distinguishability measures in
information theory, with a list of 528 references on related
topics \cite{fuchs96}.

R\'{e}nyi entropies form an important family of one-parametric
extensions of the Shannon entropy (\ref{shnent}). For
$0<\alpha\neq1$, the R\'{e}nyi entropy of order $\alpha$ is defined
as \cite{renyi61}
\begin{equation}
R_{\alpha}(X):=
\frac{1}{1-\alpha}{\ }\log\!\left(\sum\nolimits_{x}{p(x)^{\alpha}}
\right)
. \label{renent}
\end{equation}
R\'{e}nyi considered (\ref{renent}) in connection with formal
postulates characterizing entropic functions \cite{renyi61}. The
entropy (\ref{renent}) is a non-increasing function of $\alpha$
\cite{renyi61}. The entropy is maximized for the uniform
distribution: if $p(x)=1/N$ for all $x$ then (\ref{renent})
becomes $\log{N}$, or merely $1$ for $N=2$. In the limit
$\alpha\to1$, the R\'{e}nyi entropy gives (\ref{shnent}). The
joint $\alpha$-entropy $R_{\alpha}(X,Y)$ is defined in line with
(\ref{renent}) by substituting the joint probabilities.

Some special choices of $\alpha$ are widely used in the
literature. The limit $\alpha\to0$ leads to the max-entropy
equal to logarithm of the number of non-zero probabilities. In the
binary case, non-trivial max-entropies are all equal to $\log2=1$.
Taking the limit $\alpha\to\infty$, we obtain the min-entropy
$R_{\infty}(X)=-\log\max{p}(x)$. The case $\alpha=2$ gives the
so-called collision entropy, which will play an important role in
our discussion:
\begin{equation}
R_{2}(X)=-\log\!\left(\sum\nolimits_{x}{p(x)^{2}}\right)
. \label{2ent}
\end{equation}
The question of concavity should be emphasized separately. For
$\alpha\in[0;1]$, the R\'{e}nyi entropy is concave irrespectively
to the actual dimensionality \cite{ja04}. Convexity properties of
$R_{\alpha}(X)$ with orders $\alpha>1$ depend on dimensionality of
probabilistic vectors \cite{ben78,bengtsson}. For instance, the
binary R\'{e}nyi entropy is concave for $\alpha\in[0;2]$
\cite{ben78}.

To extend the notion of mutual information to the R\'{e}nyi case,
we should define the corresponding conditional form. There is no
generally accepted definition of conditional R\'{e}nyi's entropy
\cite{tma12}. Similarly to (\ref{pshen}), for $0<\alpha\neq1$ we introduce the function
\begin{equation}
R_{\alpha}(X|y):=
\frac{1}{1-\alpha}{\ }\log\!\left(\sum\nolimits_{x}{p(x|y)^{\alpha}}\right)
. \label{rect2}
\end{equation}
For $\alpha=0$, we mean here logarithm of the number of non-zero conditional probabilities.
The conditional $\alpha$-entropy is then
defined by \cite{cch97,Kam98,EP04}
\begin{equation}
R_{\alpha}(X|Y):=\sum\nolimits_{y}{p(y)\,R_{\alpha}(X|y)}
\ . \label{rect1}
\end{equation}
The limit $\alpha\to1$ gives the standard conditional entropy
(\ref{cshen}). Further, the case $\alpha=\infty$ leads to the
conditional min-entropy. For the given $y$, we define
\begin{equation}
\hat{x}(y):={\mathrm{Arg}}\>{\underset{x}{\max}}\,p(x|y)
\ , \label{htx12}
\end{equation}
so that $p(x|y)\leq{p}(\hat{x}|y)$ for all possible $x$ at the
fixed $y$. The formula (\ref{rect2}) then reads
\begin{equation}
R_{\infty}(X|y)=-\log{p}(\hat{x}|y)
\ . \label{ryin}
\end{equation}
Substituting (\ref{ryin}) into the right-hand side of
(\ref{rect1}), we get the conditional min-entropy
$R_{\infty}(X|Y)$. Similarly to (\ref{renent}), the quantity
(\ref{rect2}) is a non-increasing function of $\alpha$. Hence, we
have $R_{\infty}(X|Y)\leq{R}_{\alpha}(X|Y)$ for all $\alpha\geq0$.

We now ask for a behavior of (\ref{rect1}) under conditioning on
more. As a corollary of the concavity, for $\alpha\in[0;1]$ we
have \cite{rastit}
\begin{equation}
R_{\alpha}(X|Y,Z)\leq{R}_{\alpha}(X|Y)
\ . \label{rcoonm}
\end{equation}
The question is more difficult for $\alpha>1$. As mentioned in
section 2.3 of \cite{bengtsson}, the R\'{e}nyi entropy is not
concave for $\alpha>\alpha_{*}>1$, where $\alpha_{*}$ depends on
the number $N$ of possible outcomes. In the binary case, we can
apply (\ref{rcoonm}) for all positive orders up to $\alpha=2$.
Unfortunately, no sufficiently exact estimations of the quantity
$\alpha_{*}(N)$ are known. For more than two outcomes, we cannot
use (\ref{rcoonm}) for $\alpha=2$ and for larger values. Entropic
functions of the R\'{e}nyi type do not share some properties
satisfied by the standard entropic functions. In particular, the
conditional form (\ref{rect1}) of order $\alpha\neq1$ does not
generally obey the chain rule. A comparison of proposed forms of
conditional R\'{e}nyi's entropy is given in \cite{tma12}.

We now proceed to a discussion of the so-called ``R\'{e}nyi mutual
information''. By an analogy with the formula (\ref{dfmin}), one
introduces the quantity
\begin{equation}
I^{(R)}_{\alpha}(X,Y):=R_{\alpha}(X)-R_{\alpha}(X|Y)
\ . \label{rmidf}
\end{equation}
It could be interpreted as R\'{e}nyi's version of mutual
information. With $\alpha=2$, this quantity is widely used in
studying a performance of probes in individual attacks on quantum
cryptographic systems
\cite{brandt03q,brandt05,shapiro06,srsf98,shapiro06w}. Security
against collective attacks is typically studied with using the
Holevo information \cite{sbpc09}. Despite of analogy between
(\ref{dfmin}) and (\ref{rmidf}), the latter with $\alpha\neq1$ is
different in some essential respects. First, the quantity
(\ref{rmidf}) is not generally symmetric in its entries. Second,
the parameter $\alpha$ runs a continuum of values, so that a
proper choice of its value is not {\it a priori} clear. In a
certain sense, advantages of an approach with generalized entropic
functions is rather associated with a possibility to vary the used
parameter \cite{Kam98}. Third, the conditional $\alpha$-entropy
(\ref{rect1}) does not share the chain rule. Hence, we may fail
with interpreting (\ref{rmidf}) as a reduction in the uncertainty
of one random variable due to knowledge of other.

The above facts advise that we be very careful in treatment of the
quantity (\ref{rmidf}), especially when only one value
$\alpha\neq1$ is involved. We will further use generally accepted
term ``R\'{e}nyi's mutual information''. Due to the above reasons,
however, this quantity is not a completely legitimate measure of
mutual information. For $\alpha>1$, we can rather interpret it as
a family of upper bounds on the standard mutual information. In
the considered model of key distribution, the bits are completely
random and the scheme works symmetrically with respect to the
values ``0'' and ``1''. This results in the fact that an entropy
of each binary variable {\it per se} reaches its maximum $\log2=1$
irrespectively to $\alpha$. Let random variable $X$ with $N$
possible outcomes be distributed uniformly, i.e.,
$R_{\alpha}(X)=\log{N}$ independently of $\alpha$. For
$\alpha>\beta$, we then have
\begin{equation}
I^{(R)}_{\alpha}(X,Y)\geq{I}^{(R)}_{\beta}(X,Y)
\ . \label{mofor2}
\end{equation}
It follows from $R_{\alpha}(X|Y)\leq{R}_{\beta}(X|Y)$ for
$\alpha>\beta$, as the R\'{e}nyi entropy of $X$ is constant. Thus,
R\'{e}nyi's mutual information of order $\alpha>1$ gives an upper
bound on the standard mutual information $I(X,Y)$. Properties of
the above information measures will be essential in studies of
their adequacy. The two versions of the FPB probe form a very
spectacular example for considering this question.

\section{Two versions of the FPB probe for eavesdropping on the BB84 scheme}\label{sec3}

In this section, we recall basic details of the BB84 protocol,
which is most analyzed and most often implemented \cite{assche06}.
In this paper, we do not consider the problem of optimization of
probe characteristics. Rather, we wish to compare several possible
measures to estimate a performance of given probes. Two different
approaches can be used on Eve's side during an attack on the BB84
protocol of quantum key distribution. In any case, the
eavesdropper is asked by distinguishing a quantum state from
several known alternatives. The two basic schemes of
distinguishing non-identical pure states are referred to as the
minimum error discrimination and the unambiguous discrimination.

The problem of optimization of entangling probe for an individual
attack was considered by Fuchs and Peres \cite{fperes96}. They
showed numerically that the optimal detection method for a
two-level system can be obtained with a two-dimensional probe
only. The authors of \cite{srsf98} have made detailed analysis for
the BB84 scheme \cite{bb84}. In the paper \cite{brandt05pra},
Brandt showed that the obtained probe for attacking the BB84
scheme can be realized with a single CNOT gate. Following
\cite{shapiro06,shapiro06w}, this probe will be referred to as the
Fuchs--Peres--Brandt (FPB) probe. The analysis of \cite{srsf98} is
related to the case with the error-discard as reconciliation
procedure. When Alice and Bob use other reconciliation methods,
the probe considered is not optimal \cite{herb08}.

Let Alice and Bob use the two polarization bases
$\bigl\{|h\rangle,|v\rangle\bigr\}$ and
$\bigl\{|r\rangle,|\ell\rangle\bigr\}$. With respect to the
horizontal polarization, the kets $|r\rangle$ and $|\ell\rangle$
of the diagonal basis relate to the angles $\pi/4$ and $3\pi/4$,
respectively. In each bit interval Alice sends a single photon
prepared accordingly. Eve uses this photon as the control qubit
input to a CNOT gate. Computational basis of this gate is defined
in terms of the polarization states as
\begin{align}
|0\rangle&=\cos(\pi/8)|h\rangle+\sin(\pi/8)|v\rangle
\ , \label{0def}\\
|1\rangle&=-\sin(\pi/8)|h\rangle+\cos(\pi/8)|v\rangle
\ . \label{1def}
\end{align}
Eve prepares her own probe photon in the initial state
\begin{equation}
|t_{in}\rangle=c\,|+\rangle+s\,|-\rangle
\ , \label{tcs}
\end{equation}
where $c=\sqrt{1-2P_{E}}$, $s=\sqrt{2P_{E}}$, and
$|\pm\rangle=\bigl(|0\rangle\pm|1\rangle\bigr)/\sqrt{2}$. The
introduced parameter $P_{E}\in[0;1/2]$ turns out to be the error
probability \cite{shapiro06,shapiro06w}. The state (\ref{tcs}) is
input as the target qubit into Eve's CNOT gate controlled by sent
Alice's qubit.

To find probabilities of outcomes, the output of the gate should
be expressed as a superposition, in which the first qubit is
represented in a proper basis. It will be convenient to introduce
sub-normalized vectors
\begin{align}
|t_{\pm}\rangle&=c\,|+\rangle\pm\frac{s}{\sqrt{2}}\>|-\rangle
\ , \label{tpmdef}\\
|t_{E}\rangle&=\frac{s}{\sqrt{2}}\>|-\rangle
\ . \label{tedef}
\end{align}
When Alice uses the basis $\bigl\{|h\rangle,|v\rangle\bigr\}$,
Eve's CNOT gate acts as
\begin{align}
|h\rangle\otimes|t_{in}\rangle&\longmapsto
|h\rangle\otimes|t_{+}\rangle+|v\rangle\otimes|t_{E}\rangle
\ , \label{htin}\\
|v\rangle\otimes|t_{in}\rangle&\longmapsto
|v\rangle\otimes|t_{-}\rangle+|h\rangle\otimes|t_{E}\rangle
\ . \label{vtin}
\end{align}
In a similar manner, for the basis
$\bigl\{|r\rangle,|\ell\rangle\bigr\}$ we obtain
\begin{align}
|r\rangle\otimes|t_{in}\rangle&\longmapsto
|r\rangle\otimes|t_{+}\rangle-|\ell\rangle\otimes|t_{E}\rangle
\ , \label{m4tin}\\
|\ell\rangle\otimes|t_{in}\rangle&\longmapsto
|\ell\rangle\otimes|t_{-}\rangle-|r\rangle\otimes|t_{E}\rangle
\ . \label{p4tin}
\end{align}
Suppose that Bob applies the basis that Alice has employed and his
outcome matches what Alice sent. This case focuses on the
error-free sifted bits shared by Alice and Bob. To learn their
shared bit value, Eve should distinguish between the
sub-normalized outputs $|t_{+}\rangle$ and $|t_{-}\rangle$ of the
target qubit. At this stage, she could apply two different
approaches mentioned above.

In the Helstrom scheme of distinguishing between $|t_{+}\rangle$
and $|t_{-}\rangle$, the average error probability is minimized
\cite{helstrom67,helstrom}. In the case considered, this
measurement is described by projectors $|0\rangle\langle0|$ and
$|1\rangle\langle1|$. For any of the states $|t_{+}\rangle$ and
$|t_{-}\rangle$, the false-alarm probability is equal to
\begin{equation}
\frac{1}{2}
\left(
1-\frac{\sqrt{4P_{E}(1-2P_{E})}}{1-P_{E}}
\right)
\, . \label{hser}
\end{equation}
The second approach is known as the unambiguous discrimination
\cite{ivan87,dieks,peres1}. It sometimes gives an
inconclusive answer, but never makes an error of
mis-identification. Let us define two sub-normalized states
\begin{equation}
|\tau_{\pm}\rangle=\frac{s}{\sqrt{2}}\>|+\rangle\mp{c}\,|-\rangle
\ , \label{taudf}
\end{equation}
so that
$\langle\tau_{+}|t_{+}\rangle=\langle\tau_{-}|t_{-}\rangle=0$. The
POVM elements of unambiguous discrimination are then written as
\cite{shapiro06}
\begin{equation}
\mm_{+}=\frac{1}{2c^{2}}\>|\tau_{-}\rangle\langle\tau_{-}|
\ , \qquad
\mm_{-}=\frac{1}{2c^{2}}\>|\tau_{+}\rangle\langle\tau_{+}|
\ , \qquad
\mm_{?}=\frac{2c^{2}-s^{2}}{2c^{2}}\>|+\rangle\langle+|
\ . \label{mmpmin}
\end{equation}
The probability of inconclusive answer is equal to the overlap of
the states to be discriminated \cite{peres1}. Calculating the
overlap with the normalized vectors, we get
\begin{equation}
\frac{c^{2}-s^{2}/2}{c^{2}+s^{2}/2}=\frac{1-3P_{E}}{1-P_{E}}
\ . \label{pina}
\end{equation}
We shall now obtain values of any mutual information as a function
of $P_{E}$.

To analyze a probe performance, calculations will be
conditioned on the error-free sifted bits shared by Alice and Bob
\cite{shapiro06,shapiro06w}. Let us begin with the case, when Eve
uses the Helstrom scheme. By the prime sign, we will mean terms
related to the error-free sifted bits. It is also assumed that
$0\leq{P}_{E}\leq1/3$. The formula (\ref{hser}) merely gives the
conditional probabilities $p(e^{\prime}=0|b^{\prime}=1)$ and
$p(e^{\prime}=1|b^{\prime}=0)$. As Alice's bits are equally likely
to be $0$ and $1$, two possible values of $b^{\prime}$ are equally
likely as well. Multiplying (\ref{hser}) by $1/2$, we then obtain
the joint probabilities $p(e^{\prime},b^{\prime})$ for
$e^{\prime}\neq{b}^{\prime}$. The joint probabilities for
the case $e^{\prime}=b^{\prime}$ read
\begin{equation}
p(e^{\prime}=j,b^{\prime}=j)=\frac{1+\varkappa}{4}
\ , \qquad
\varkappa=\frac{\sqrt{4P_{E}(1-2P_{E})}}{1-P_{E}}
\ , \label{hser1}
\end{equation}
where $j=0,1$. It is also clear that here we have the symmetry
$p(e^{\prime}|b^{\prime})=p(b^{\prime}|e^{\prime})$. With these
facts, we easily obtain any quantity representing an amount of
mutual information. The standard mutual information is written as
\begin{equation}
I(B^{\prime},E^{\prime})=
\frac{1+\varkappa}{2}\>\log(1+\varkappa)+\frac{1-\varkappa}{2}\>\log(1-\varkappa)
\ . \label{fpsm}
\end{equation}
The R\'{e}nyi mutual information of order $\alpha=2$ becomes
\cite{shapiro06,shapiro06w}
\begin{equation}
I_{2}^{(R)}(B^{\prime},E^{\prime})=
\log\bigl(1+\varkappa^{2}\bigr)
\ . \label{fprm2}
\end{equation}
It is typically used for estimating a performance of
quantum cryptographic probes
\cite{brandt03q,brandt05,shapiro06,srsf98,shapiro06w}. At the same
time, there are no fundamental reasons to prefer just the order
$\alpha=2$. Advantages of a consideration with generalized
entropies are rather connected with variations of entropic
parameters. We will also include the case $\alpha=\infty$ due to
(\ref{mofor2}). For the first version of the FPB probe, we have
\begin{equation}
I_{\infty}^{(R)}(B^{\prime},E^{\prime})=
\log(1+\varkappa)
\ . \label{fprm8}
\end{equation}
This value shows the least point of the interval, in which the
mutual $\alpha$-information ranges for the given $P_{E}$.
Comparing the three quantities (\ref{fpsm}), (\ref{fprm2}), and
(\ref{fprm8}) with the case of conclusive probe will show 
deficiencies of using (\ref{rmidf}) as a measure of Eve's information gain. The
quantities (\ref{fpsm}), (\ref{fprm2}), and (\ref{fprm8}) are
symmetric with respect to the entries $B^{\prime}$ and
$E^{\prime}$. It is not the case for the probe with unambiguous
discrimination.

In the conclusive modification, Eve discriminates between
$|t_{+}\rangle$ and $|t_{-}\rangle$ unambiguously. This case is
more complicated as involving an inconclusive answer. 
Restricted to the error-free sifted bits of Alice and Bob, the
probabilities of outcomes are
\begin{equation}
p(e^{\prime}=0)=p(e^{\prime}=1)=\frac{P_{E}}{1-P_{E}}
\ , \qquad
p(e^{\prime}=?)=\frac{1-3P_{E}}{1-P_{E}}
\ . \label{peps}
\end{equation}
We further have $p(b^{\prime}=j|e^{\prime}=j)=1$ and
$p(b^{\prime}=j|e^{\prime}=?)=1/2$ for $j=0,1$ \cite{shapiro06}.
For the conclusive probe, the standard mutual information is equal
to
\begin{equation}
\widetilde{I}(B^{\prime},E^{\prime})=
1-\frac{1-3P_{E}}{1-P_{E}}=
\frac{2P_{E}}{1-P_{E}}
\ . \label{fpbsm}
\end{equation}
For all $\alpha\geq0$, the R\'{e}nyi mutual $\alpha$-information
is also given by the right-hand side of (\ref{fpbsm}). That is, it
does not depend on $\alpha$. This property reflects that the
situation is completely deterministic whenever $e^{\prime}\neq?$
too. Hence, for all $\alpha\geq0$ we have
\begin{equation}
R_{\alpha}(B^{\prime}|e^{\prime}=0)=R_{\alpha}(B^{\prime}|e^{\prime}=1)=0
\ . \label{r01eq0}
\end{equation}
For $e^{\prime}=?$, two possible values of $B^{\prime}$ are
equiprobable, so that
$R_{\alpha}(B^{\prime}|e^{\prime}=?)=\log2=1$. For all
$\alpha\geq0$, therefore, the conditional entropy reads
\begin{equation}
R_{\alpha}(B^{\prime}|E^{\prime})=
p(e^{\prime}=?)\,R_{\alpha}(B^{\prime}|e^{\prime}=?)=\frac{1-3P_{E}}{1-P_{E}}
\ . \label{rafix}
\end{equation}
Bob's bits are equally likely to be $0$ and $1$, whence
$R_{\alpha}(B^{\prime})=\log2=1$. Combining the latter with
(\ref{rafix}), we actually get
\begin{equation}
\widetilde{I}_{\alpha}^{(R)}(B^{\prime},E^{\prime})=
\frac{2P_{E}}{1-P_{E}}
\ . \label{fpbrm}
\end{equation}
Thus, the R\'{e}nyi mutual information increases with $\alpha$ for
the usual probe and is constant for the conclusive probe. The
above results will be used in analysis of an adequacy of
R\'{e}nyi's mutual information in the context of quantum
cryptography.

\section{R\'{e}nyi's mutual information as a measure of probe performance}\label{sec4}

In this section, we consider quantities of the
form (\ref{rmidf}) as estimators of a performance of quantum
cryptographic probes. In particular, we will focus on the
R\'{e}nyi mutual information of order $\alpha=2$. The author of
\cite{shapiro06} examined the conclusive version of the FPB probe
just on the ground of this measure. On the other hand, there is no
fundamental reasons to prefer namely the order $\alpha=2$. Such
choice of entropic parameter was rather used in \cite{srsf98} for
simplifying analytical calculations. Indeed, explicit formulas of
optimization problem seem to be most tractable for $\alpha=2$ too.
However, some convenience of calculations hardly is a proper
motivation to choose the basic figure of a probe performance. The standard mutual information
(\ref{mindf}) obey many nice properties that justify its usual
interpretation. Such properties are connected with characteristics
of the conditional entropy (\ref{cshen}). The definition (\ref{rect1}) of conditional R\'{e}nyi's entropy is
losing many of desired properties. In particular, it does not share the
usual chain rule. Only for two outcomes we can be sure that
conditioning reduces the entropy with orders up to $2$. For $\alpha>1$, the R\'{e}nyi entropy is neither
purely concave nor purely convex. The answer crucially depends on the number
of possible outcomes \cite{ben78}. In general, the R\'{e}nyi
mutual information is not symmetric in its entries. For the
conclusive modification of the FPB probe, we generally have
\begin{equation}
\widetilde{I}_{\alpha}^{(R)}(B^{\prime},E^{\prime})\neq
\widetilde{I}_{\alpha}^{(R)}(E^{\prime},B^{\prime})
\qquad (\alpha\neq1)
\ . \label{nonsym}
\end{equation}
The left-hand side of (\ref{nonsym}) is always given by
(\ref{fpbrm}), whereas the right-hand side of (\ref{nonsym})
essentially depends on $\alpha$. Note that for the usual FPB probe
we actually have the symmetry
$I_{\alpha}^{(R)}(B^{\prime},E^{\prime})=I_{\alpha}^{(R)}(E^{\prime},B^{\prime})$.
This result follows from the fact
$p(b^{\prime}|e^{\prime})=p(e^{\prime}|b^{\prime})$ already
mentioned for the probe with Helstrom's scheme of discrimination.
In general, however, such symmetry is not valid.

The above reasons show difficulties with the quantity (\ref{rmidf}). At the
same time, these reasons are rather general in character. To give a
complete picture, we must mention one fact proved by the authors
of \cite{srsf98}. For the original formulation of the B92 scheme,
they proved that the standard mutual information and R\'{e}nyi's
information of order $2$ are maximized by the same optimal probe.
It may be not valid for sophisticated protocols or scenarios.
Moreover, this property {\it per se} does not justify that
R\'{e}nyi's mutual information of order $2$ is always able to
replace the standard one. We shall now exemplify some
inadequacy of (\ref{rmidf}) in the context of quantum cryptography. 
This conclusion is easily demonstrated by
comparing the two versions of the FPB probe from the viewpoint
of measures of the form (\ref{rmidf}) with several $\alpha$.
Apparently, it is better represented by a contrast between the
curves, which show various measures as functions of $P_{E}$. As
usual, we will focus on the interval $0\leq{P}_{E}\leq1/3$.

\begin{figure}
\includegraphics{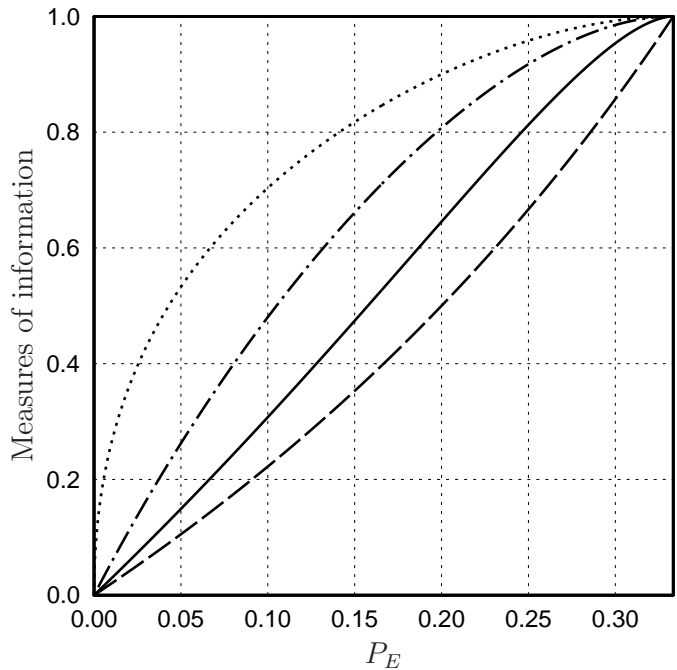}
\caption{For the FPB probe with projective
measurement, it shows (\ref{fpsm}) by {\it solid line}, (\ref{fprm2}) by
{\it dash-dotted line}, and (\ref{fprm8}) by {\it dotted line}. For the
conclusive probe, it shows (\ref{fpbrm}) by {\it dashed line}.}
\label{fig1}
\end{figure}

On Fig. \ref{fig1}, the four curves are plotted. For the FPB probe
with Helstrom's scheme, one shows (\ref{fpsm}) by solid line,
(\ref{fprm2}) by dash-dotted line, and (\ref{fprm8}) by dotted
line. For the conclusive FPB probe, we show all the measures of
mutual information by dashed line. Including the quantity
(\ref{fprm8}) with $\alpha=\infty$ allows to describe an interval,
in which the R\'{e}nyi mutual information may range with changing
$\alpha$. In the paper \cite{shapiro06}, the two
probes were compared on the base of R\'{e}nyi's mutual information
of order $2$. As we see from Fig. \ref{fig1}, a difference between
dash-dotted and dashed lines is sufficiently large and reaches the
maximum about $0.314$. On this ground, a performance of the
conclusive probe seems to be weak too, even for small values of
$P_{E}$. However, this difference can be changed by varying
$\alpha$, since the quantity (\ref{fpbrm}) is independent of
$\alpha$ due to a specific character of the conclusive probe. If
we choose $\alpha=\infty$, then a difference is even more
essential and reaches the maximum about $0.482$.

However, our conclusions on security of cryptographic schemes
should be independent of a sporadic choice. Keeping in mind the
result (\ref{mofor2}), we conclude that R\'{e}nyi's information of
order $\alpha>1$ should rather be treated as an upper bound on the
mutual information. It is clearly demonstrated by the solid,
dotted and dash-dotted lines for the FPB probe. The standard
measure (\ref{mindf}) of mutual information obeys many properties
including the relation (\ref{dfmin}). It is for this reason that
it is to be treated as a reduction in the uncertainty of one
random variable due to knowledge of other. On the other hand,
there is no known analog of the result (\ref{dfmin}) with measures
of the R\'{e}nyi type. Moreover, we have arrived at the fact
(\ref{nonsym}). To make a valid decision on Eve's mutual
information, the standard mutual information is most appropriate.
This conclusion is brightly illustrated by the curves of Fig.
\ref{fig1}. Comparing the solid and dashed lines shows that the
conclusive probe is certainly not so weak, as it could be imagined
with the dash-dotted line. Despite of some non-optimality, the
conclusive probe may nevertheless be of interest due to additional
reasons.

It is instructive to look at the situation in slightly other
perspective. As Bob's binary variable is uniformly distributed, we
have $R_{\alpha}(B^{\prime})=\log2=1$ irrespectively to $\alpha$.
Hence, the maximization of Eve's mutual information is
equivalent to the minimization of the corresponding conditional
entropy. In this regard, we see a similarity with one scenario of
entropic uncertainty relations for successive measurement
\cite{bfs2014,zzhang14}. We already mentioned that the definition
(\ref{rect1}) loses some properties that are natural for a measure
of conditional entropy. Since the term
$R_{\alpha}(B^{\prime}|E^{\prime})$ does not increase with
$\alpha$, for $\alpha>1$ it gives a lower bound on the conditional
entropy $H(B^{\prime}|E^{\prime})$. This point is a counterpart of
the fact that R\'{e}nyi's mutual information of order $\alpha>1$
leads to an upper bound on the standard one.

Using the measures (\ref{fprm2}) or (\ref{fprm8}), the legitimate
users tend to be sure in an essential weakness of the conclusive
version in comparison with the usual FPB probe. So, they could
underestimate the degree of vulnerability with respect to
conclusive eavesdropping. On the other hand, an overstated
estimation of probe power can lead to a wrong situation with
practical realization of a communication system. As was remarked
in section VI.L of \cite{tittel}, the infinite security implies
the infinite cost. Developing this observation, we may arrive at a
conclusion. Characteristics of a real cryptographic system are
actually determined by the taken compromise between several
conflicting requirements such as an amount of provided security
versus paid costs or technological tools. In this regard, we wish
to estimate precisely a performance of possible probes.

As curves of Fig. \ref{fig1} show, using R\'{e}nyi's mutual
information may create an illusory ``power'' in amount of Eve's
mutual information. A degree of this illusion depends on the
choice of $\alpha>1$. Such a wrong expression about too strong
probe performance may lead to rejecting some feasible
realizations. They could seemingly be evaluated as unexpectedly
vulnerable, even though they are suitable in other respects. For
small values of $P_{E}$, the ratio of (\ref{fprm2}) to
(\ref{fpsm}) can approach $2$ arbitrarily close. Hence, a quality
of the usual FPB probe in comparison with the conclusive one may
be illusory up to two times. The mentioned difference is most
essential for ``weak'' eavesdropping, when Eve attempts to get a
small amount of information while causing only a slight
disturbance. Nevertheless, this example illustrates a conclusion
that security requirement could be overstated spuriously on the
base of inappropriate measures of mutual information.

In principle, we may try to adopt the right-hand side of
(\ref{nonsym}), which is the $\alpha$-entropy of Eve's bits minus
the $\alpha$-entropy conditioned on Bob's bits. Our example shows,
however, that such measures of mutual information are completely
inappropriate. For the conclusive version, we obtain the
conditional entropies
\begin{equation}
p(e^{\prime}=j|b^{\prime}=j)=\frac{2P_{E}}{1-P_{E}}
\ , \qquad
p(e^{\prime}=?|b^{\prime}=j)=\frac{1-3P_{E}}{1-P_{E}}
\ . \label{cpeb}
\end{equation}
where $j=0,1$. To illustrate an inadequacy of
$\widetilde{I}_{\alpha}^{(R)}(E^{\prime},B^{\prime})$ for
$\alpha\neq1$, we calculate it for the least order
$\alpha=\infty$. Using (\ref{peps}) and (\ref{cpeb}), we finally
obtain
\begin{equation}
\widetilde{I}_{\infty}^{(R)}(E^{\prime},B^{\prime})=
\begin{cases}
0& \text{if }\,  0\leq{P}_{E}\leq1/5\,, \\
\log(2P_{E})-\log(1-3P_{E})& \text{if }\, 1/5<P_{E}<1/4\,, \\
1& \text{if }\, 1/4\leq{P}_{E}\leq1/3\,.
\end{cases}
\label{irinf}
\end{equation}
That is, this quantity sharply increases from $0$ to $1$ in a
narrow interval of $P_{E}$. For $\alpha=2$, we also get the
quantity
\begin{equation}
\widetilde{I}_{2}^{(R)}(E^{\prime},B^{\prime})=
\log\!\left(
\frac{4P_{E}^{2}+(1-3P_{E})^{2}}{2P_{E}^{2}+(1-3P_{E})^{2}}
\right)
\ . \label{iren2}
\end{equation}
The latter presents slightly smoothed variant of (\ref{irinf})
and also has a narrow interval of sharp increasing. This behavior
quite differs from (\ref{fpbsm}). Thus, the measures (\ref{irinf})
and (\ref{iren2}) hardly represent anything with respect to 
the conclusive probe.

In the context of quantum cryptography, unambiguous state
discrimination is also interesting in another respect. The rate of
inconclusive tests is actually an additional check parameter. This
fact was actually noticed in \cite{palma} with respect to the B92
scheme with conclusive receiver. For some of individual
entangling-probe attacks against this setup, Eve's activity
affects a rate of inconclusive outcomes. As was mentioned in
\cite{palma}, unchanged rate of inconclusive results cannot be
treated as a security confirmation. On the other hand, unexpected
variations of this rate are probably a witness of opposite
activity. The same fact seems to be valid with respect to Eve's
party. Observing surprise changes in inconclusive tests, Eve will
rather decide that her opponents have made counter measures. Of
course, any party may influence on a rate of inconclusive outcomes
by means of the so-called fake states. However, such an approach
will demand additional costs and more complicated equipment. From
this viewpoint, the use of unambiguous discrimination is certainly
of practical interest.

The following question is also related to conclusive
eavesdropping. As was discussed in section 4 of \cite{shapiro06},
conclusive eavesdropping is especially interesting when there is
loss between Alice's transmitter and Bob's receiver. Here, Eve can
act as follows. She captures each carrier from Alice and
recognizes it using the unambiguous state discrimination. Then she
prepares qubit for transmission to Bob, whenever the test has
given the conclusive answer. When the probability of conclusive
result is higher than the transmissivity of the communication
channel, the quantum cryptography seems to be inappropriately
vulnerable. In the terminology of the paper \cite{palma}, this
strategy corresponds to opaque eavesdropping. It seems that opaque
eavesdropping on the BB84 scheme has received less attention than
it deserves. In any case, studies of such a kind should use 
properly chosen quantifier of mutual information.

\section{Conclusions}\label{sec5}

Using the two versions of eavesdropping with the FPB probe, we
compared different quantifiers of an amount of Eve's mutual
information about the error-free sifted bits. The so-called
R\'{e}nyi mutual information, usually used with order $2$, is not
a completely legitimate measure of the mutual information. For
$\alpha>1$, it should rather be interpreted as an upper bound on
the standard mutual information. The standard mutual information
must be considered as a preferable measure, at least in studies of
conclusive eavesdropping. These conclusions were expected from a
consideration of properties of related information-theoretic
quantities. Further, we explicitly demonstrate our findings by
comparison of the conclusive modification with the usual one. The 
use of inappropriate measures of mutual
information could lead to wrong decisions in practical building a
communication with quantum key distribution.

Finally, we shall briefly discuss possible alternatives to using
information functions of the R\'{e}nyi type. Tsallis' entropies
\cite{tsallis} form another especially important family of
generalized entropies. Two conditional forms of the Tsallis
entropy were considered \cite{sf06}. First of them obeys the chain
rule of usual form, whence Tsallis' counterpart of the mutual
information is arisen \cite{sf06}. In principle, such measures of
information could be applied in the context of eavesdropping.
Another way is to use information metrics of this type. The standard
conditional entropy leads to an information metric between random
variables \cite{benn98}. Each of two conditional forms of Tsallis'
entropy also provides a legitimate metric \cite{rastann}, though
the second one does not share the chain rule. In Section
\ref{sec4}, we already noticed that maximizing mutual information
reached with the FPB probe can be rewritten as minimizing the
corresponding conditional entropy. So, it is of interest to study
a probe performance from the viewpoint of information metrics.

\end{document}